# Magnetodielectric Coupling in Nonmagnetic Au/GaAs:Si Schottky Barriers

S. Tongay[1] and A. F. Hebard[1]

[1]*Department of Physics, University of Florida, Gainesville, FL 32611-8440*

Y. Hikita[2] and H. Y. Hwang[2,3]

[2]*Department of Advanced Materials Science, University of Tokyo, Kashiwa, Chiba 277-8561, Japan*

[3]*Japan Science and Technology Agency, Kawaguchi, Saitama 332-0012, Japan.*

We report on a heretofore unnoted giant negative magnetocapacitance (>20%) in non-magnetic Au/GaAs:Si Schottky barriers that we attribute to a magnetic field induced increase in the binding energy of the shallow donor Si impurity atoms. Depletion capacitance ($C_{dep}$) dispersion identifies the impurity ionization and capture processes that give rise to a magnetic field dependent density of ionized impurities. Internal photoemission experiments confirm that the large field-induced shifts in the built-in potential, inferred from $1/C_{\mathrm{dep}}^2$ vs voltage measurements, are not due to a field-dependent Schottky barrier height, thus requiring a modification of the abrupt junction approximation that accounts for the observed magnetodielectric coupling.

PACS: 73.30.+y, 71.20.Nr, 71.55.Eq

## (I) Introduction

Electronic transport across metal-semiconductor interfaces, which are ubiquitous in semiconductor technology, is mediated by the formation of Schottky barriers and associated depletion capacitance. Numerous studies have established the relationship between the Schottky barrier height and materials properties and have at the same time fully characterized the dependence of electronic transport across metal-semiconductor interfaces on temperature, frequency and voltage bias/polarity[1-8]. Despite decades of investigations and the use of concepts such as metal induced gap state models[2],[3] and bond polarization theory[4,5], a consensus understanding of Schottky barriers has not been reached. Schottky contacts on the semiconductor GaAs are particularly interesting due to considerations such as a long spin life time in GaAs[9], the demonstration of spin polarized current injection from a metal into GaAs[10] and spin extraction from GaAs into a metal[11]. An additional gap in knowledge however becomes apparent with the realization that there are relatively few studies of the effect of externally applied magnetic fields *H* on the electrical properties of Schottky barriers.

We address this deficiency by reporting on a surprisingly large negative magnetocapacitance (MC > 20%) associated with Au/GaAs:Si Schottky barrier samples fabricated and characterized by standard techniques as described below. The MC is independent of field direction and is unexpected because (1) there are no magnetic impurities in the Au/GaAs:Si system and (2) the GaAs:Si is homogeneous and thus not a candidate for “magnetocapacitance in nonmagnetic composite media”[12]. Clearly, such large MC effects in non-magnetic semiconductor systems must be explained before the be-havior of metal-semiconductor interfaces involving spin-polarized metals and/or dilute magnetic semiconductors (DMS) for spintronics applications[13] can be understood. Using a combination of current-voltage, capacitance ($C$) and internal photoemission (IPE) studies, we show that the MC can be attributed to a novel magnetodielectric coupling in which a $H$-induced increase in the binding energy of the Si donor impurities strongly affects the density of ionized impurities ($N_d$) within the depletion width of the Schottky barrier, and hence the polarization. We identify the ionization and capture transitions between the shallow impurity $E_{sh}$ and conduction $E_c$ bands (see schematic of Fig. 2 inset) and show that the apparent large $H$-induced increase of the built-in potential $V_{bi}$ deduced from linear $1/C_{dep}^2$ versus reverse-voltage bias $V_R$ plots is not due to a change in the Schottky barrier height $\Phi_{SBH}$, but rather to a field induced change in the binding energy, $E_a = E_c - E_{sh}$, of the Si donor impurities. Magnetic field tunability of the Schottky capacitance offers a new degree of freedom in the design and application of magneto electronic devices.

## (II) Experimental Methods

Commercially available GaAs wafers with a nominal Si dopant density of $3\times10^{16}$ cm$^{-3}$ were used. Ohmic contacts were made based on multilayer recipes existing in the literature[14-16]. Rapid thermal anneals in nitrogen gas at temperatures in the range 400°C - 460°C assured good ohmic contact with low parasitic resistance down to temperatures as low as 10 K. Prior to evaporating the front-side Au Schottky contact, the samples were thoroughly cleaned in 3:1:50 $HNO_3$:HF:$H_2O$ for 3-4 minutes to remove any native oxide. Nine separate samples prepared with different contacts were studied, all giving similar results. At room temperature the experimental values for $N_d$ determined from $1/C^2$ vs $V$ measurements were found to be in the range $1\text{-}2\times10^{16}$ cm$^{-3}$.

The current voltage (I-V) characteristics of a high quality Schottky diode should exhibit pronounced asymmetry with respect to the sign of the bias voltage and also be well described in forward bias by thermionic emission[1]. These attributes are satisfied for our samples as shown in Fig. 1a by the forward and reverse bias I-V characteristics, measured at 300K and 20 K, of the same Au/GaAs:Si Schottky sample on which ac impedance measurements are performed. At 300 K the Schottky barrier height, $\Phi_{SBH}$, and the ideality factor, η, are extracted for a total of nine different samples using thermionic emission theory[1] and found to be respectively in the ranges 0.82 V < $\Phi_{SBH}$ < 0.92 V and 1.05 < η < 1.15. These values are in good agreement with the current literature[1] and with our IPE data discussed below. At low temperature (20 K) the I-V characteristics of the same sample (Fig. 1b, blue diamonds) show similar high quality rectifying behavior; reliable barrier parameters however cannot be extracted using thermionic emission theory because of competing processes such as generation-recombination, quantum tunneling and thermionic-field emission[1, 6, 17] taking place at low temperatures.

Complex impedance measurements over a frequency range 20Hz-1MHz were made using an HP4284 capacitance bridge. The output of the bridge can be interpreted by one of two different models each having two unknowns: a resistance $R_s$ in series with a capacitance $C_s$ (series model) or a resistance $R_p$ in parallel with a capacitance $C_p$ (parallel model). The simplest model for a Schottky diode however involves at least three unknowns: $R_\Sigma$ in series with a two-component complex capacitance $C^*$ which comprises the capacitance $C = \mathrm{Re}\{C^*\}$ of the depletion region in parallel with a loss term that represents dc transport (tunneling) and ac loss due to changes in polarization. It is straightforward to show that $C$ is bounded by $C_s$ and $C_p$ (i.e., $C_s > C > C_p$) and that if $C_s$ is found to be close to $C_p$, then $R_\Sigma$ is small and can be ignored[8]. For the measurements reported here we find *at most* $C_s = 1.07C_p$, thus implying the narrow constraint $1.07C_p > C > C_p$, and hence the assurance that the measured $C_p$ of the parallel model accurately represents the depletion capacitance $C_{\mathrm{dep}}$. Since $R_\Sigma$ is negligible, then any field dependence of $R_\Sigma$ is also negligible, and we can conclude that the measured MC is associated with the Schottky depletion capacitance rather than magnetoresistance in the contacts or bulk masquerading as magnetocapacitance[18]. At temperatures lower than the 20 K, a rapid increase in $R_\Sigma$ manifests itself as a large difference in $C_p$ and $C_s$ (i.e., $C_s >> C_p$), and $C_p$ is no longer an accurate measure of the depletion capacitance[8]

We have also used internal photoemission (IPE) to directly measure $\Phi_{\mathrm{SBH}}$ . The sample is mounted in a cryostat and illuminated by ac modulated light emerging from an optical fiber[19]. The internal photocurrent through the sample is synchronously demodulated and the square root of the photoyield defined as the photocurrent per incident photon, is plotted against photon energy. A linear extrapolation to zero gives the minimum energy, i.e. $\Phi_{\mathrm{SBH}}$, necessary to excite electrons from the Fermi energy of the metal over the barrier.

**(III) Experimental Results**

In Fig. 2 we show as our main result the effect of magnetic field on capacitance. The magnitude of the relative change in capacitance $\Delta C_{\mathrm{dep}}/C_{\mathrm{dep}}$ measured at frequency $f$ = 1 MHz and over a field range of 0-70 kOe is observed to increase as the temperature $T$ is lowered from 300 K to 20 K. Near freeze-out temperatures[20] the MC grows rapidly, reaching -21% at $H$ = 70 kOe and $T$ = 20 K. To understand these data, we generalize the Mott-Schottky (M-S) picture[1] by explicitly including the independent variables, $f$, $T$ and $H$, and writing,

$$\frac{1}{C_{dep}(f,T,H)^2} = \frac{2(V_{bi}(f,T,H)+V_R)}{e\varepsilon_s N_d(f,T,H)} \quad , \quad W(f,T,H) = \left[\frac{2\varepsilon_s(V_{bi}(f,T,H)+V_R)}{eN_d(f,T,H)}\right]^{1/2} \quad , \qquad (1)$$

where $\varepsilon_S$ is the dielectric constant of the semiconductor, $V_R$ is the magnitude of the applied reverse bias voltage (metal electrode is negative), and $N_d(f,T,H)$ is the density of ionized Si impurities within the depletion width $W(f,T,H)$. In agreement with Eq. (1), a linear dependence of $1/C_{\mathrm{dep}}^2$ with respect to $V_R$ is found at $f$ = 1 MHz for different $H$ (Fig. 3, top panel) and at $H$ = 0 for different $f$ (Fig. 3, bottom panel). For each data set there are two extracted parameters: the slope, from which $N_d(f,T,H)$ can be calculated (Eq. 1), and the intercept $V_{bi}(f,T,H)$, the built-in potential. At room temperature almost all the donor electrons are excited into the conduction band leaving the donor atoms fully ionized with a density $N_d^0(T=300\ K) \approx N_d$. However, temperature and magnetic field freeze-out take place at lower temperatures, and $N_d(f,T,H)$ becomes a function of $H$ at fixed $T$ as shown in the inset of the lower panel of Fig. 3. The related changes of $V_{bi}(f,T,H)$ are also shown in the same inset. We note that $N_d(f,T,H)$ and $V_{bi}(f,T,H)$ extracted from the linear plots of Fig. 3 can be used in Eq. 1 to calculate the MC (e.g., squares in Fig. 2 for $T$ = 20 K) and are found to be in good self-consistent agreement for all measured $f$, $T$ and $H$. The remainder of this paper will focus on elucidating the mag-netodielectric coupling that gives rise to the pronounced $f$, $T$ and $H$ dependence of the extracted M-S parameters of Eq. 1 and the associated MC shown in Fig. 2.

The underlying physical processes are revealed in the frequency-dependent capacitance and loss plots of Fig. 4. There are two prominent loss peak regions: the first low-frequency region extends over the frequency range 100 Hz – 10 kHz and the second high-frequency region, with more pronounced loss, extends from 10 kHz to greater than the 1 MHz limit of our capacitance bridge. With decreasing $T$ and/or increasing $H$, the lossy regions move to lower frequency as shown in the successive panels of Fig. 4.

We find that each loss curve is well described by the imaginary part of the ubiquitous Cole-Davidson expression[21] for the generalized dielectric constant,

$$\varepsilon = \varepsilon_\infty + \frac{\varepsilon_0 - \varepsilon_\infty}{1 + (i\omega\tau)^{1-\beta}} \quad , \qquad (2)$$

where $\varepsilon_\infty$ and $\varepsilon_0$ are the dielectric constants in the high and low frequency limits, $\beta$ is a constant smaller than one, $\tau$ is a relaxation time and $\omega$ the angular frequency. The analysis of the low-frequency peaks is straightforward, since the frequency range is broad enough to include the full peak, thereby enabling us to determine in a straightforward manner the peak frequency $f_p = 1/\tau$ and a corresponding relaxation time $\tau$ at which the loss for a particular $T$ and $H$ peaks at a maximum. For constant $H$ the relaxation rate $\tau^{-1}$ adheres to a thermally-activated Arrhenius dependence, $\tau^{-1} = \tau_0^{-1}\exp(-E_a/k_BT)$, where $E_a$ is an activation energy and $\tau_0^{-1}$ a prefactor. The semilogarithmic plot of $\tau^{-1}$ versus $1/T$ for the low-frequency peak shown in the inset of Fig. 5 ($H$ = 70 kOe) manifests typical activated response. The field-dependent activation energies $E_a(H)$ are extracted from the slopes of these lines and plotted in the main panel of Fig. 5 against field (red squares) for the low-frequency loss peaks.

Since the amplitude and shape of the low-frequency loss peaks remain constant and are shifted only laterally with temperature/field (see Fig. 4), we can extract similar activation energies for the high-frequency peaks by monitoring the temperature dependence of an arbitrary point (e.g., half amplitude) rather than the peak. By making the reasonable assumption that the high-frequency peaks also have an invariant amplitude and shape, we extract the activation energies shown as blue circles in Fig. 5. The error bars on these data are larger because the high frequency portions of these peaks are greater than 1 MHz, and there is consequently greater uncertainty in the parameters of the Cole-Davidson fits.

The loss peaks of Fig. 4 correspond to two relaxation processes, each having similar activation energies which increase with magnetic field (Fig. 5). At $H = 0$, $E_a = 6.05 \pm 0.20$ meV and $5.73 \pm 0.13$ meV for the low and high peaks respectively, very close to the reported values of 5.8 meV[1,22] for Si impurities in GaAs. We therefore attribute the two observed loss peaks to ionization from $E_{sh}$ to $E_c$ of the dopant valence electrons and capture from $E_c$ back to $E_{sh}$. The separate relaxation frequencies imply that the rate for ionization $\gamma_{sh\to c}$ is different than the rate $\gamma_{c\to sh}$ for capture. Using detailed balance[23], i.e., $\gamma_{sh\to c} n_{sh} = \gamma_{c\to sh} n_c$, where $n_{sh}$ and $n_c$ represent respectively the density of carriers in $E_{sh}$ and $E_c$, we can infer $\gamma_{c\to sh} > \gamma_{sh\to c}$, since $n_{sh} > n_c$ near freeze out where the MC is dominant. This argument allows us to identify the low frequency loss peak $\gamma_{sh \to c}$ with ionization and the high frequency peak $\gamma_{c \to sh}$ with capture. Intraband transitions such as hopping are not seen because our measurement temperatures are too high.

The increase in $E_a$ with respect to $H$ has been previously studied both theoretically[24,25] and experimentally using optical[26,27] and Hall measurements[22,28]. In the GaAs host, the Bohr radii of the hydrogen-like donor electrons are renormalized upwards by the small effective mass, $m_{GaAs} = 0.065m_e$, and the large relative permittivity, $\varepsilon_{GaAs} = 13.5$. The resulting magnetic freeze-out brings electrons closer to the donor atoms, thus increasing the Coulomb energy and $E_a$. The renormalization by the host lattice sharply reduces the fields well below the levels required to see an observable effect for hydrogen atoms in vacuum. Our observed change in $E_a$ with $H$ is in good qualitative agreement in functional form[28] and magnitude[25,22] with previous experiments on Si-doped GaAs.

The above interpretation suggests that at low $T$ electrons can be frozen out from the conduction band $E_c$ to the impurity band donors as $H$ increases. This capture process explains the $H$-induced decrease in $N_d(f,T,H)$ at fixed $f$ and $T$ shown in the lower panel of Fig. 3. It does not explain however the $H$-dependence of $V_{bi}(f,T,H)$. Usually, $V_{bi}$ extracted from a M-S analysis is used to calculate $\Phi_{SBH}$ from the relation[1], $\Phi_{SBH} = V_{bi} + (E_c - E_F)/e$. An increase in $V_{bi}$ corresponds by Eq. 1 to an increase in $W$ and a corresponding decrease in $C_{dep}$. Since $E_c$-$E_F$, which is calculated to be ~10 meV, is a small correction with negligible $H$ dependence, the measured shift in $V_{bi}$ of 300 meV for a 7 T change in $H$ (see Fig. 3 top panel) implies that there is a comparable shift in $\Phi_{SBH}$. Such a dependence of $\Phi_{SBH}$ on $H$ extracted from $C$ measurements is unphysical since, as shown by the schematic in the inset of Fig. 2, the capacitance, $C_{bond}$, arising from bond polarization and the associated dipole layer giving rise to $\Phi_{\mathrm{SBH}}$ at the metal-semiconductor interface is in series with the much smaller $C_{\mathrm{dep}}$ and hence can be ignored. The expected $H$ independence was checked using IPE measurements[19, 29, 30] on similar samples at various temperatures and fields up to 70 kOe. The intercepts of the linearly extrapolated photocurrent yield on the abscissa of the inset to the top panel of Fig. 3 show a small $T$ dependence[6] but no $H$ dependence. Thus the $H$-induced $V_{bi}$ shift has another origin and the M-S equations must be modified.

The above conclusion also applies to the magnetic field dependence of the forward-biased threshold observed in the I-V characteristics. In Fig. 1b we show the I-V characteristics at the four indicated temperatures for two values of field, $H$ = 0 Oe (solid symbols) and $H$ = 70 kOe (open symbols). We note that while the magnetic field has negligible effect on the forward bias threshold at high temperatures (300 K), there is a systematic increase in the field-dependent increment of the forward bias threshold as $T$ is lowered to 20K. At 20 K the ideality factor is significantly larger than unity, thus indicating that accurate estimates of $\Phi_{\mathrm{SBH}}$ cannot be obtained due to the presence of additional processes such as generation-recombination, quantum tunneling and thermionic-field emission[1, 6, 17]. The observed field-induced changes in the low temperature forwardbiased thresholds are thus more likely attributed to corrections to $\Phi_{\mathrm{SBH}}$ derived from image-lowering and thermal field emission. These two processes become important at temperatures much lower than the apparent barrier height and give a correction that decreases when $N_D(f,H,T)$ decreases[31, 32]. Thus with increasing $H$, the concomitant decrease in $N_D(f,H,T)$ leads to an increase in the "effective" $\Phi_{\mathrm{SBH}}$ and the forward bias threshold as seen in Fig. 1b. However, our calculations show that these corrections are only responsible for ~10 meV increase in barrier height and thus do not explain the large shift in built-in potential seen in Fig. 3. We conclude that the interpretation of low temperature I-V curves using traditional thermionic emission theory, which remains an open problem in the literature[6], is unreliable. For these reasons we focus on ac impedance measurements, which identify the frequency magnetic field dependent processes occurring within the depletion region, and the internal photoemission (IPE) measurements, which directly determine $\Phi_{SBH}$.

**(IV) Modification of the abrupt junction approximation**

The M-S relations for $C_{dep}$ are usually derived using the abrupt junction approximation (AJA), phenomenological description, which assumes a constant density $N_d$ of ionized impurities within the depletion region in which $N_d(f,T,H) = N_d^0(T)$ is constant for all $x$ within the depletion width ($0 \leq x \leq W$) and zero elsewhere. Although the AJA is overly simplistic, it has the well-known advantage that when combined with Poisson's equation the relationship between $1/C_{dep}^2$ and reverse bias voltage $V_R$ is linear. This linearity of the Mott-Schottky (M-S) plots is seen in many experiments[1] including ours (Fig. 3). The failure of the AJA to include frequency (and field) dependence is an obvious deficiency. Thus, if $C_{dep}$ is frequency (magnetic field) dependent as it is in our experiment (Fig. 3), the extracted slopes and intercepts are by necessity also frequency (magnetic field) dependent (Fig. 3) and cannot be simply related to the high temperature (fully ionized) values of the built-in potential $V_{bi}^0$ and ionized impurity density $N_d^0$. Our solution to this deficiency, a deficiency which has been recognized in previous work[7, 33, 34], is the introduction of a modification of the AJA as discussed below.

As *T* is lowered, electrons from the conduction band are captured and $N_d^0(T)$ decreases. To incorporate the effects of *f* and *H*, we modify the AJA with the expression,

$$N_d(x, f, T, H) = N_d^0(T) + N_{cap}(f, T, H)\left[\Theta(y-x)\exp(-x/L) + \alpha\right] + N_{ion}(f, T, H)\exp(x/W)\ , \quad (3)$$

where $N_{\text{cap}}$ and $N_{\text{ion}}$ represent respectively the additional charge (and hence polarization) of the high-frequency capture and low-frequency ionization processes, $\Theta$ is the unit step function, *L* is a characteristic length and *y*, which obeys the constraint $0 < y < W$, is a cutoff beyond which the exponential contribution to $N_{\text{cap}}$ is zero. The constant $\alpha$ assures that capture processes can occur for all *x*. Following Bauza[7], who offers a similar expression to account for deep level electronic traps of Ti-W/*n*-Si Schottky diodes, we solve Poisson's equation using Eq. 3 and find,

$$\frac{1}{C_{dep}^2} = \frac{2\left(V_{bi}^0(T) - eN_{cap}(f,T,H)L^2\left(e^{(-y/L)}\left(1+y/L\right)\right)/\varepsilon_s + V_R\right)}{e\varepsilon_s\left(N_d^0(T) + \alpha N_{cap}(f,T,H) + 2N_{ion}(f,T,H)\right)} \quad (4)$$

Although the phenomenology is considerably more complex with the introduction of five additional unknown parameters ($N_{cap}$, $N_{ion}$, $L$, $y$, $\alpha$), our formulation has the distinct advantage that linearity of the M-S plots is preserved. The understanding gained from this reformulation is the redefinition of the extracted slopes and intercepts, i.e.,

$$N_d(f,T,H) = N_d^0(T) + \alpha N_{cap}(f,T,H) + 2N_{ion}(f,T,H)\,, \quad (5)$$

and

$$V_{bi}\left(f,T,H\right)=V_{bi}^{0}(T)-eN_{cap}(f,T,H)L^{2}\left(e^{(-y/L)}\left(1+y/L\right)\right)/\varepsilon_{s} \quad . \qquad (6)$$

In the high $T$ limit where all impurities are ionized, i.e., $N_{\text{cap}} \rightarrow 0$ and $N_{\text{ion}} \rightarrow 0$, Eq. 4 reduces to the conventional $f$ and $H$-independent M-S relation, $1/C_{dep}^{2}(T)=2(V_{bi}^{0}(T)+V_{R})/e\varepsilon_{s}N_{d}^{0}(T)$ and Schottky-Mott relation is preserved. With decreasing $T$, $V_{bi}^{0}$(T) (and hence $\Phi_{\text{SBH}}$) increases due to temperature variations of the bandgap[6], and the decrease in $N_d^0(T)$ (thermal freezeout) is compensated by increases in $N_{\text{cap}}$ and $N_{\text{ion}}$, reflecting the dominance of $f$- and $H$-dependent processes.

**(V) Discussion and summary**

An intuitive physical understanding of the above equations derives from the fact that $N_d$ is exponentially dependent on the ratio $-E_a(H)/k_BT$. Thus as $T$ decreases, or $H$ increases with a corresponding increase in $E_a$ (Fig. 5), $N_d^0$ decreases from its high-temperature (100% ionization) value, and there is a concomitant increase in the number of transitions between the shallow $E_{sh}$ and conduction $E_c$ bands (see arrows in the Fig. 2 schematic). At fixed $f$ and $H$ the parameters $N_{cap}$ and $N_{ion}$, which as modifications to the AJA represent the ionized donors participating in the interband transitions, thus increase with decreasing temperature. However, the relative increase of the quantity $\alpha N_{cap}(f,T,H)+2N_{ion}(f,T,H)$ in Eq. 5 with decreasing temperature is not sufficient to compensate for the decrease in $N_d^0$, thereby giving rise to a net decrease in $N_d(f,T,H)$ consistent with the decreased capacitance with decreasing temperature. At fixed $T$, the interband transitions associated with $N_{\text{cap}}$ and $N_{\text{ion}}$ and shown in Fig. 4 decrease with increasing $H$ (field freezeout) and increasing $f$ (polarization cannot follow rapid changes), thus accounting for the observed decrease in $N_d$ described by Eq. 5 and shown in the inset to the lower panel of Fig. 3.

We emphasize that Eqs. 3 and 4 represent a useful phenomenological generalization of the original M-S relations which in turn are based on the over-simplified phenomenological AJA. Importantly, for the purposes of this paper, we see from Eqs. 4 and 6 that the observed $V_{bi}$ dependence on $H$ is a function of the unknown parameter $y$, which in the limit, $y >> L$, describes capture of carriers in close proximity to the metal-semiconductor interface ($x = 0$) where electrons from the metal are readily available and in the opposite limit, $y << L$ , describes capture processes occurring uniformly throughout the depletion width.

Similar arguments apply to the observed increases in $V_{bi}$ shown for $T = 20$ K in the inset to the lower panel of Fig. 3 and described by Eq. 6. The apparent large change of $V_{bi}$ with field does not imply a corresponding change in $\Phi_{SBH}$ as some might incorrectly conclude. We attribute all of the field dependence describing our observed magneto-dielectric coupling to the magnetic field dependent donor impurity binding energy, $E_a(H)$, which in turn imposes field dependence on $N_{cap}$ and $N_{ion}$. More explicitly, at fixed magnetic field and low temperature (e.g., $T = 20$ K) where freeze-out is important and the corrections $N_{\mathrm{cap}}$ and $N_{\mathrm{ion}}$ to the AJA are significant, the extrapolated intercept $V_{bi}(f,T,H)$ described by Eq. 6 increases as the measurement frequency is swept from 100Hz to 1MHz (Fig. 3, bottom inset). At low frequencies the dominance of capture and ionization processes leads to a correction to $V_{bi}^{0}(T)$. In this low-frequency limit, $V_{bi}$ determined by M-S extrapolations agrees with $\Phi_{SBH}$ extracted from IPE measurements (Fig. 3, top in-set). Since IPE measurements directly measure $\Phi_{SBH}$ in the dc limit, it is this limit where both relaxation processes occurring within the depletion width have to be included. We have shown that the AJA approximation alone is not sufficient in predicting $\Phi_{SBH}$ and one has to take into account the frequency and field-dependent correction terms associated with the capture/ionization processes between the impurity and valence bands. As frequency increases, $V_{bi}$ also increases to significantly higher values due to the decrease in the correctional terms, $N_{\mathrm{cap}}$ and $N_{\mathrm{ion}}$, to $V_{bi}^{0}(T)$.

At the highest frequencies, beyond the 1 MHz upper bound of our measurements, $N_{\mathrm{cap}}$ and $N_{\mathrm{ion}}$ both approach zero, and the well-known AJA form, $1/C_{dep}^{2}(T)=2(V_{bi}^{0}(T)+V_{R})/e\varepsilon_{s}N_{d}^{0}(T)$, of the Mott-Schottky relation (Eq. 1) ensues. In this limit $V_{bi}$ is not a physically meaningful measure of $\Phi_{SBH}$, since as shown in this paper the overly simplistic AJA does not take into account the frequency and field dependent processes revealed so explicitly in Figs. 3 and 4. This understanding is relevant for higher frequency applications.

In summary, we have shown that magnetic freezeout[24] of shallow band carriers is responsible for significant changes in polarization within the depletion width of conventional Au/GaAs:Si Schottky barriers. Our unequivocal determination of a $H$-independent Schottky barrier height together with our identification of thermally-activated interband ionization and capture processes justifies our modification of the AJA to preserve the linearity of the $1/C_{\mathrm{dep}}^{2}$ vs $V_R$ plots (as experimentally observed) and at the same time imposes a dependence of $N_d$ and $V_{bi}$ on $f$, $T$ and $H$. Finally, the underlying magnetodielectric coupling not only allows a new experimental technique for the tuning of the do-pant carrier density at the same interface by magnetic field, but should also be important for engineering and understanding the behavior of related interfacial structures incorporating DMS and spin-polarized metals.

**Acknowledgements:** Work supported by NSF grant No. 0704240. We thank J. Nesbitt and A. Donius for experimental help at the beginning stages of this project and A. Armstrong, B. Gila, N. Newman, S. Pearton, and C. Stanton for stimulating and useful discussions.

## Figure legends:

**Figure 1:** Current-voltage characteristics show a temperature and field dependent forward bias threshold. The top panel shows the rectifying (diode) characteristics on a semi logarithmic current-voltage scale for forward (+) and reverse (-) bias voltages at 300 K (red circles) and 20 K (blue squares). The bottom panel shows on a linear current-voltage scale the forward bias onsets of conduction at the indicated temperatures for *H* = 0 (closed symbols) and *H* = 70 kOe (open symbols).

**Figure 2**: Capacitance of a Au/GaAs:Si Schottky junction decreases with increasing applied magnetic field *H*. The relative change in capacitance $\Delta C_{\mathrm{dep}}/C_{\mathrm{dep}}$ becomes increasingly more negative with increasing field. The solid curves represent data taken at the decreasing temperatures (top to bottom) indicated in the legend. The solid orange squares associated with the 20 K isotherm are calculated using Eq. 1 with the M-S parameters extracted from $1/C^2$ versus voltage plots such as shown in Fig. 3. Inset, Schematic of band bending with parameters defined in the text. Capture processes (red arrow) tend to dominate near the interface whereas ionization processes are distributed over the depletion width W. The large capacitance associated with a high density of polarized bonds at the interface is in series with the much smaller depletion capacitance.

**Figure 3:** Linearity of $1/C^2$ vs $V_R$ plots enable determination of M-S parameters. Data at 20 K are shown at 1 MHz for the indicated fields *H* (top panel) and at *H* = 0 for the indicated frequencies *f* (bottom panel). The vertical arrows mark selected extrapolated intercepts with the abscissa corresponding to the built-in potential $V_{bi}(f,H)$ at 20 K. The inset to the lower panel shows the frequency dependence of $V_{bi}$ (right hand axis) for *H* = 0 (70 kOe) for solid (open) triangles and the frequency dependence of $N_D$ (left hand axis, see text) for *H* = 0 (70 kOe) for solid (open) squares. The inset of the upper panel shows IPE data in which *H*-independent extrapolations (dotted lines) to the abscissa at the indicated temperatures imply that there is no dependence of the Schottky barrier height on *H* up to 70 kOe.

**Figure 4:** Temperature and field dependent frequency dispersion of the capacitance reveals two separate loss processes. In the three panels the frequency-dependent capacitance (left hand axis, top curves) and loss (right hand axis, bottom curves) are shown as a function of frequency for each of the three indicated temperatures. Each group of curves contains data for the five different fields indicated in the legend of the middle panel. For fixed fields the loss peaks shift to lower frequency as the temperature is reduced. The horizontal arrow in each panel marks the isothermal shift to lower frequency of the low-frequency loss peak (ionization) as the field is increased from 0 to 70 kOe.

**Figure 5:** Activated binding energy $E_a$ of Si impurity donated electrons increases with increasing field. The dependence of $E_a$ on $H$ is shown for both the low frequency (red squares, ionization) and high frequency (blue circles, capture) loss peaks. Each point is determined by fixing the field and plotting the frequencies of the loss peaks versus $1/T$ on a semilogarithmic plot. The slopes of the resulting linear plots (e.g., inset for $H$ = 70 kOe) determine the $H$-dependent binding energies.

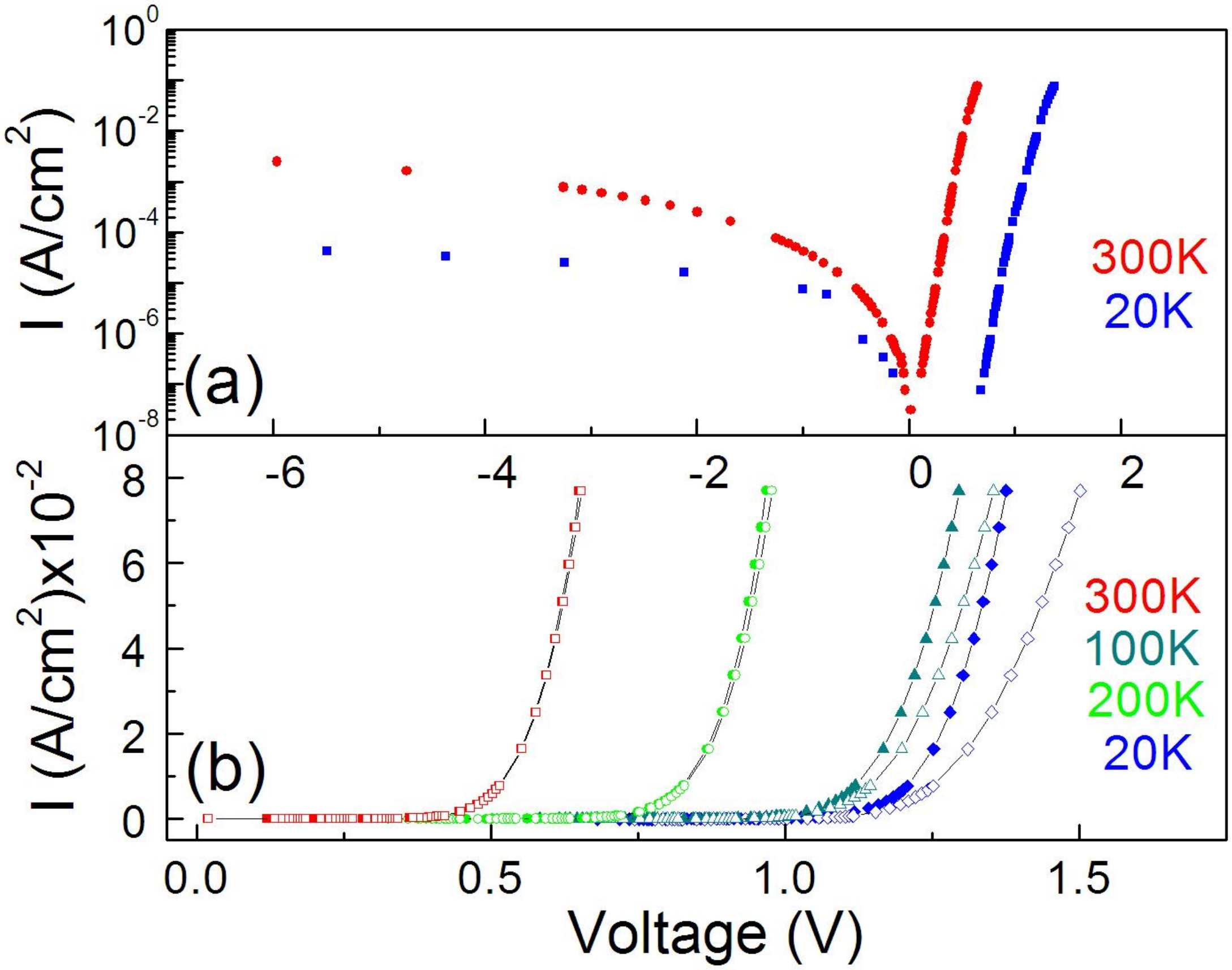

Figure 1

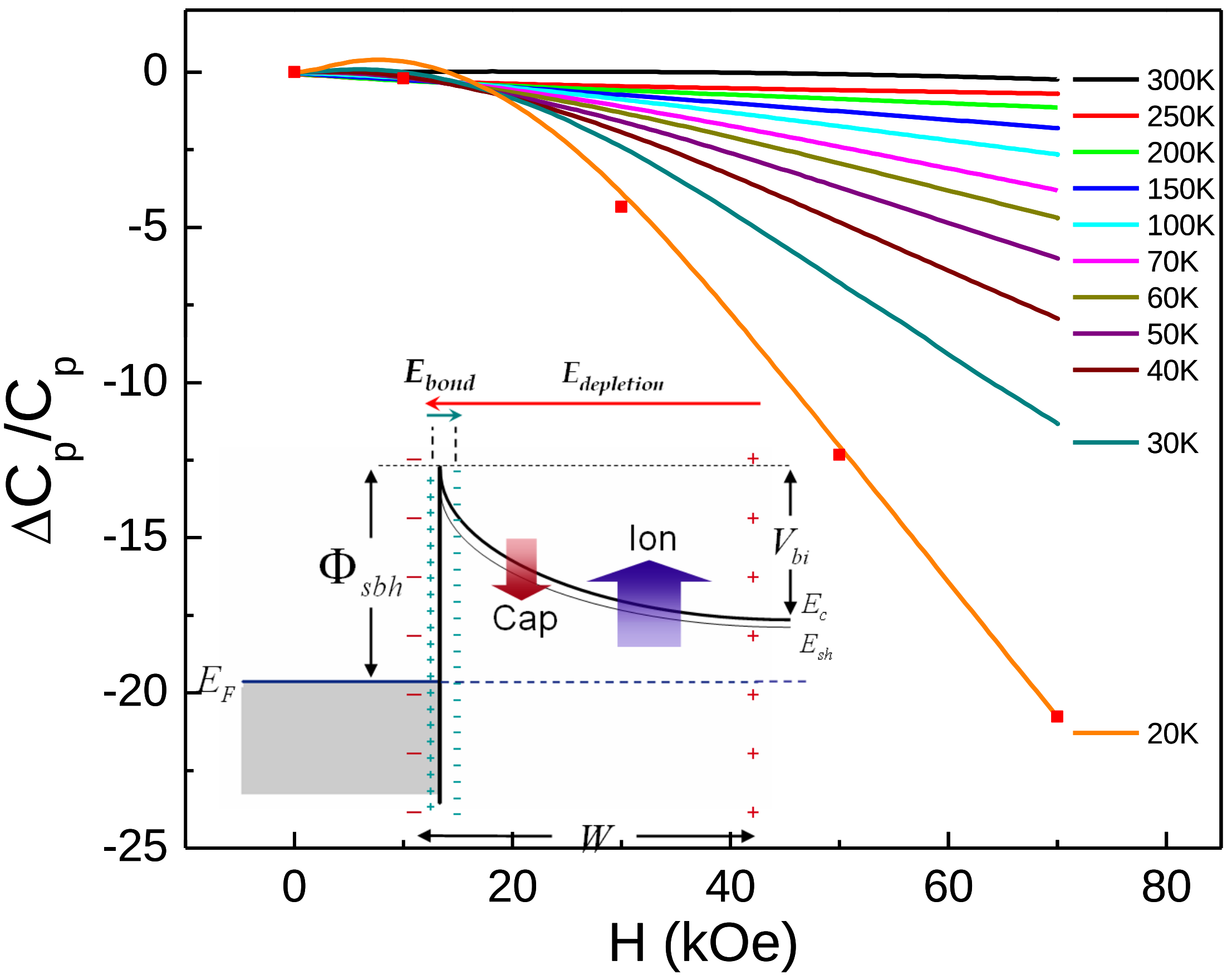

Figure 2

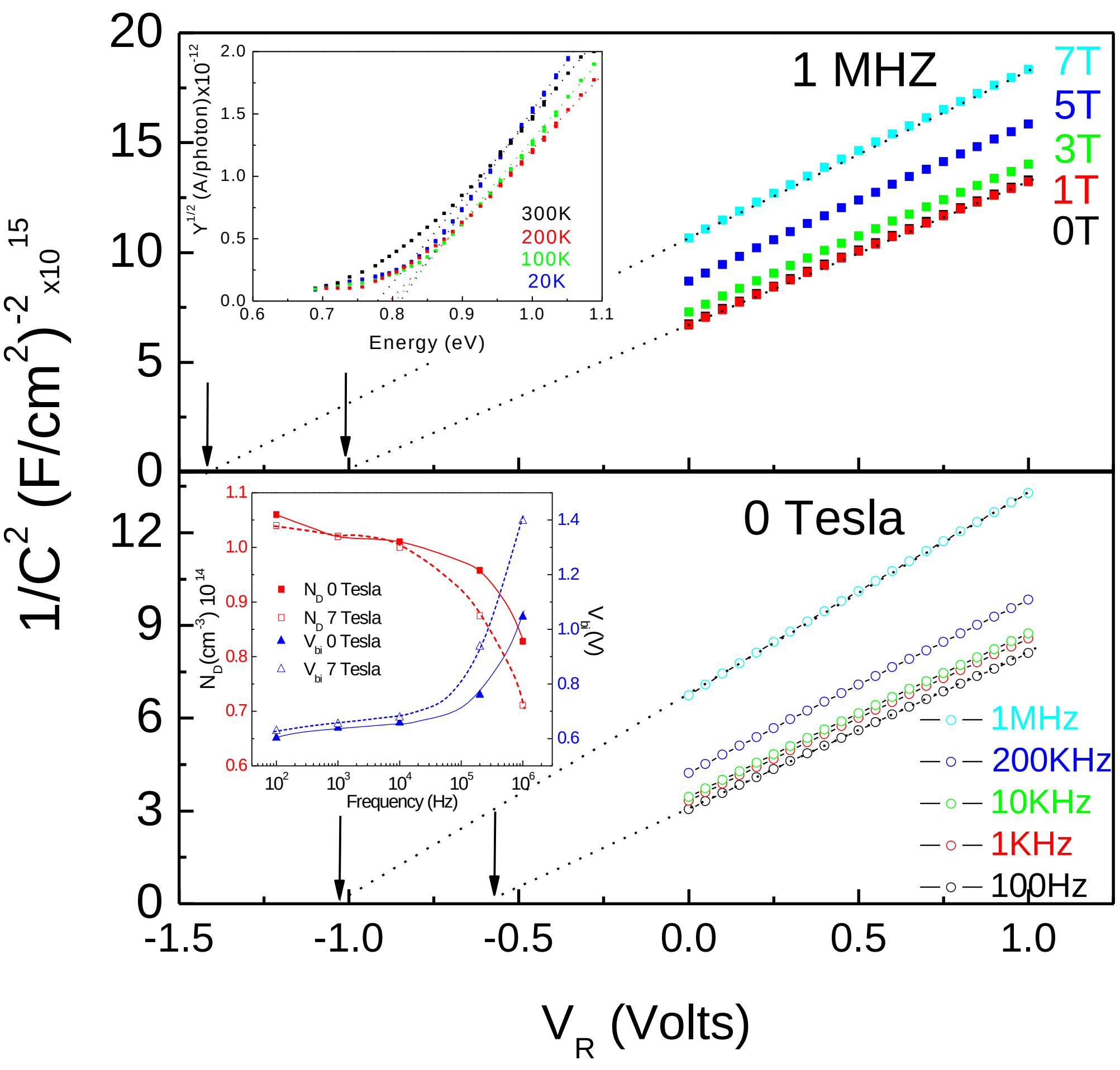

Figure 3

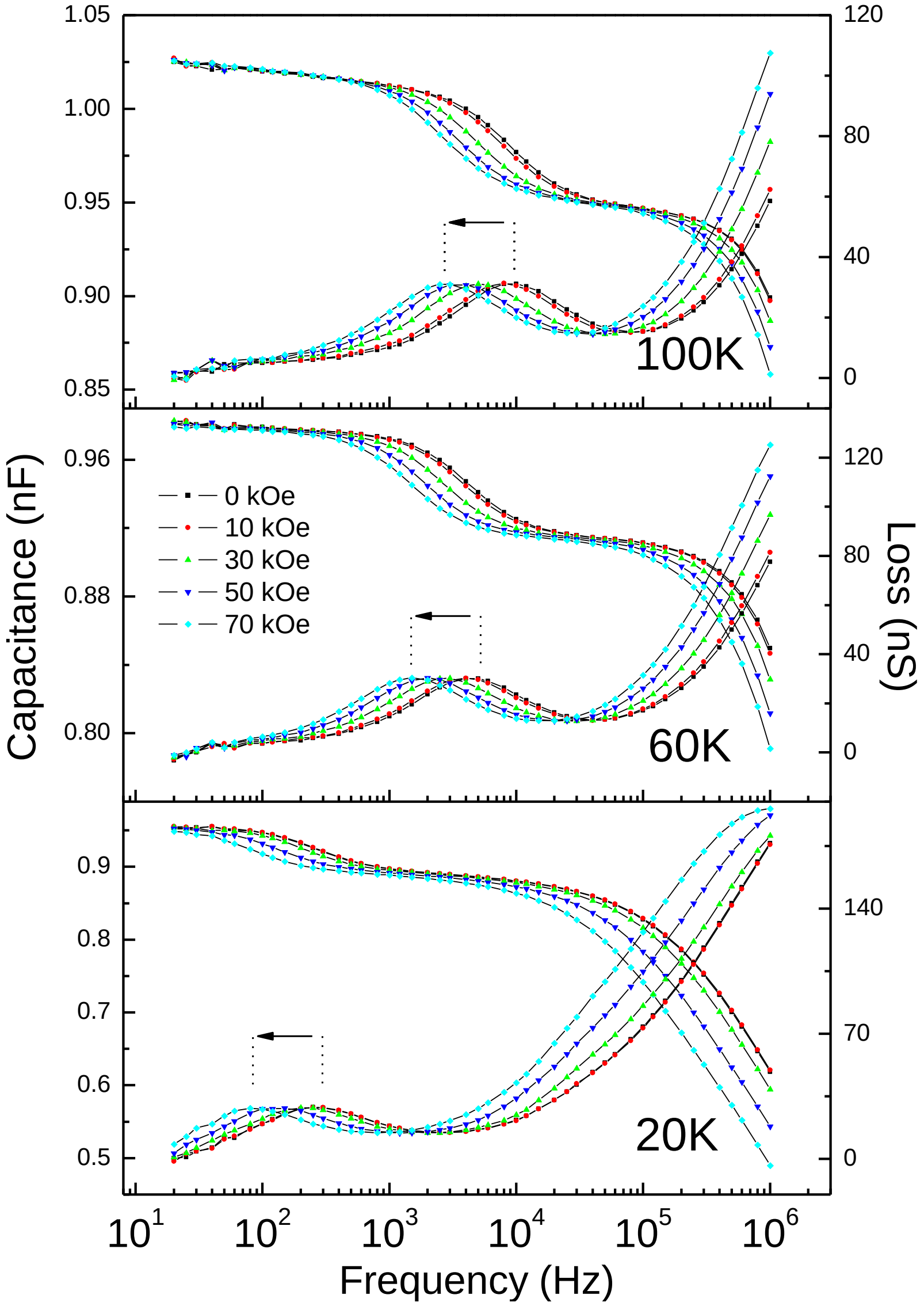

Figure 4

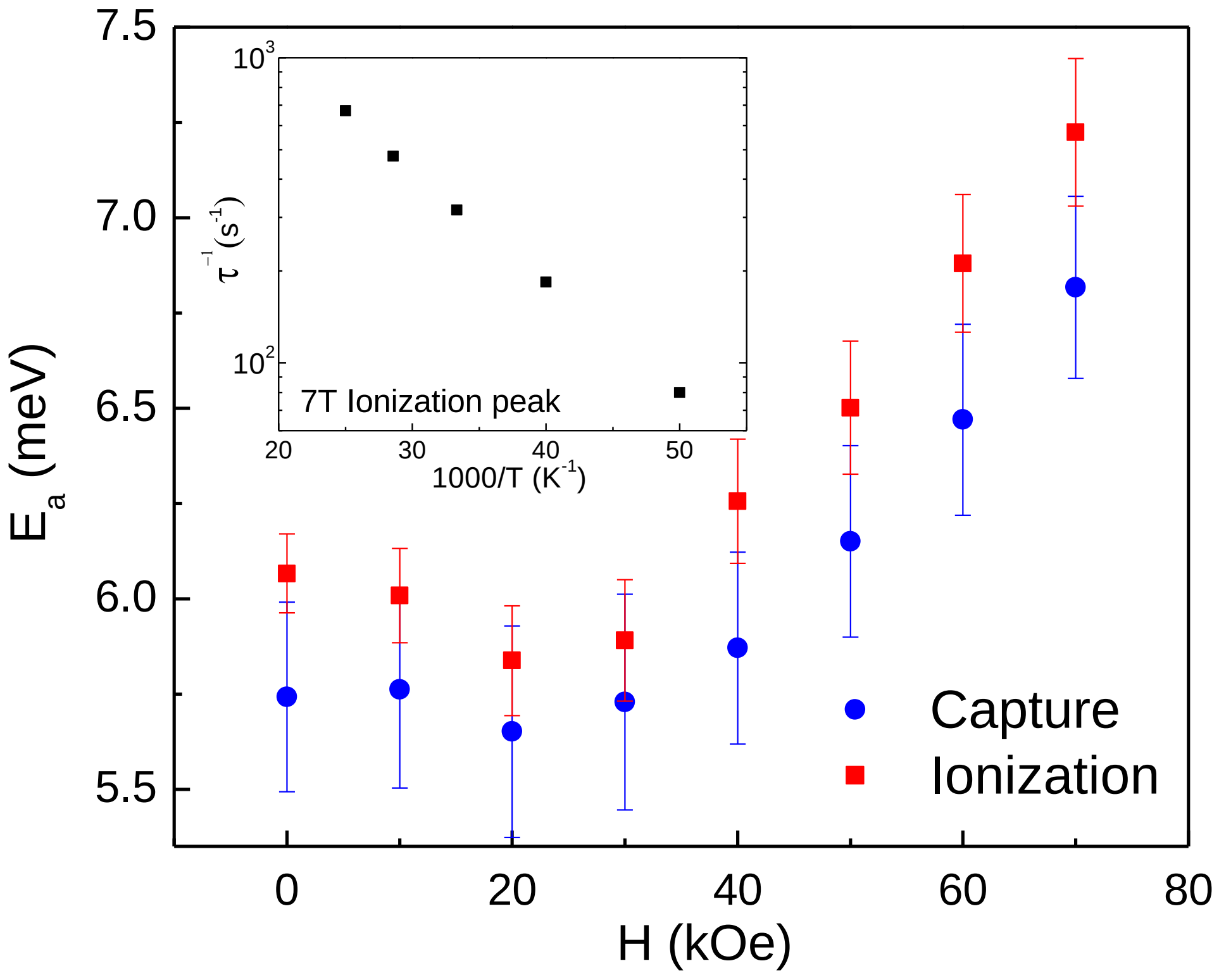

Figure 5